\documentclass[10pt, a4paper]{article}

\usepackage[T1]{fontenc}
\usepackage[utf8]{inputenc}
\usepackage[english]{babel}

\usepackage{mathptmx}
\usepackage{microtype}

\usepackage{amsmath, amssymb, amsthm}

\usepackage{booktabs}
\usepackage{multirow}
\usepackage{graphicx}
\usepackage{float}
\usepackage{subcaption}
\usepackage{array}
\usepackage{tabularx}

\usepackage{geometry}
\usepackage{multicol}           
\usepackage{balance}            
\usepackage{xcolor}
\usepackage{titlesec}           
\usepackage{abstract}           
\usepackage{fancyhdr}
\usepackage{hyperref}
\usepackage{natbib}

\titleformat{\section}{\normalsize\bfseries\scshape}{\thesection.}{0.5em}{}
\titleformat{\subsection}{\normalsize\bfseries}{\thesubsection}{0.5em}{}
\titleformat{\subsubsection}{\normalsize\itshape}{\thesubsubsection}{0.5em}{}
\titlespacing*{\section}{0pt}{8pt}{4pt}
\titlespacing*{\subsection}{0pt}{6pt}{3pt}

\hypersetup{colorlinks=true, linkcolor=blue!60!black,
            citecolor=blue!60!black, urlcolor=blue!60!black}

\begin{document}

\begin{titlepage}
  \thispagestyle{empty}
  \centering
  \vspace*{2cm}

  {\large ETH Zürich \\ Department of Management, Technology, and Economics \\[0.4em]
   Semester Project, Spring 2026}

  \vspace{2.5cm}

  \rule{\linewidth}{1pt}\\[0.6em]
{\LARGE \bfseries
The Geometry of Alliances: Vote Transfer Modelling in French Two-Round Elections\\[0.4em]
}
  \rule{\linewidth}{1pt}

  \vspace{2cm}

  {\large Emmanuel Omont}\\[0.3em]
  {\normalsize \texttt{eomont@ethz.ch}}\\[1em]
  {\normalsize Supervised by Prof.\ Barton Lee}

  \vspace{2cm}

  {\normalsize April 2026}

  \vspace{3cm}

  \begin{minipage}{0.82\linewidth}
    \small\itshape
    \textbf{Abstract.}
    Two-round legislative elections are decided not only by first-round vote shares, but by how voters whose preferred party did not advance redistribute their votes among the surviving candidates. We develop a principled model of this transfer process, grounded in multi-dimensional ideological embeddings derived from the Chapel Hill Expert Survey, and apply it to French legislative elections. Calibrated on the 2017 and 2022 elections, the model is evaluated on the unusually complex 2024 contest, in which three ideologically distinct blocs reached the second round simultaneously, achieving 90.34\% constituency-level accuracy. We show that ideological proximity alone explains the large majority of vote transfers, that a simple left--right axis is insufficient to capture the relevant distances, and that the structural configuration of the 2024 electorate was far less favourable to the far right than pre-election forecasts suggested.
    
  \end{minipage}

  \vfill
  \small
\end{titlepage}

\newpage
\thispagestyle{empty}
\tableofcontents
\newpage

\setlength{\columnsep}{0.6cm}
\begin{multicols*}{2}

\section{Introduction}
\label{sec:intro}

Two-round legislative electoral systems are common around the world. In such systems, a country is divided into constituencies; candidates compete locally; those who exceed a first-round threshold advance to a second round held shortly after, at which point voters must cast their ballot among the qualifying candidates, and the plurality winner takes the seat. Systems of this kind exist across many democracies (France, Sri Lanka, Canada...) and share a structural feature that makes them analytically rich and practically difficult to forecast: the outcome is not determined by first-round totals alone, but by the transfer of votes from eliminated parties to those who survive.

This vote transfer problem is the central subject of the present report. When a party fails to advance, its supporters must decide, or simply not show up. Their votes, in aggregate, migrate toward the surviving candidates in proportions that depend on ideological proximity \citep{downs1957}, strategic calculation, and local context. A well-documented manifestation of this dynamic is the \emph{front républicain}, a coordination mechanism whereby voters tactically support a candidate far from their own preferences in order to block a party they consider dangerous, a form of strategic electoral coordination theorised by \citet{cox1997}. That voters do systematically adapt their behaviour between rounds has been confirmed experimentally by \citet{vanderstraeten2010}, who show that subjects facing two-round rules respond to qualifying thresholds with strategic vote choices consistent with this dynamic. Understanding vote transfers is therefore not merely a forecasting exercise: it sheds light on how ideology structures coalition formation, how multi-dimensional political spaces collapse into binary or ternary contests, and how electoral rules shape the strategic behaviour of parties and voters alike. This process often look difficult to predict, but in this project, we will see how we managed to obtain a model with above 90\% accuracy prediction for the second round of the 2024 elections in France, based on the results of the 1st round.

\begin{figure}[H]
  \centering
  \includegraphics[width=\linewidth]{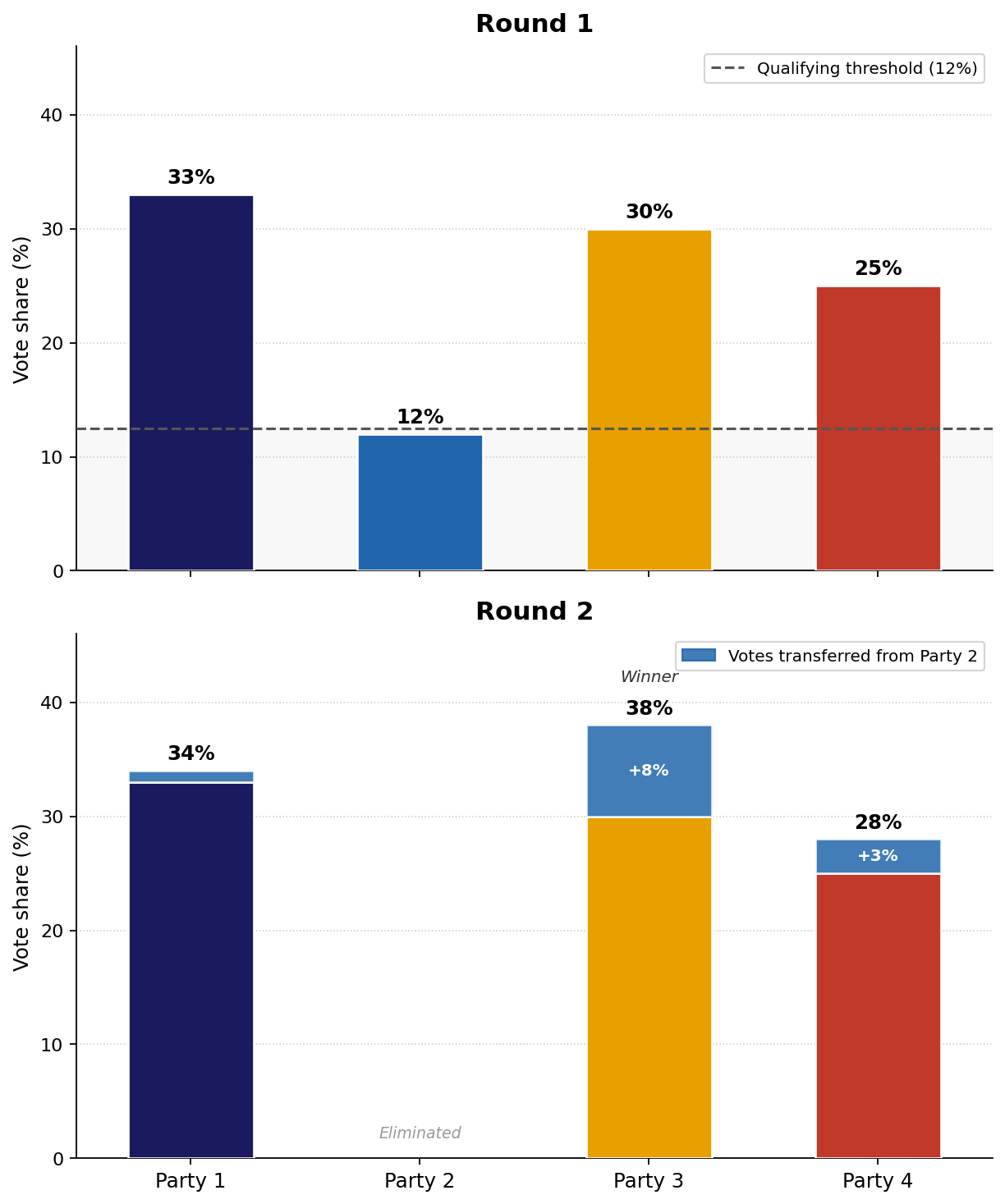}
  \caption{Schematic vote transfer in a three-candidate second round}
  \label{fig:transfer-example}
\end{figure}

The 2024 French legislative elections provide an unusually sharp test case. Following European Parliament elections in which President Macron's Renaissance party finished a distant second behind the Rassemblement National (31.4\% vs. 14.6\%), Macron dissolved the National Assembly and called early elections, a decision which caught most parties off guard. For the first time in the Fifth Republic, three large and ideologically distinct blocs simultaneously qualified for the second round across a substantial number of constituencies: the Union of the Left (Nouveau Front Populaire), encompassing La France Insoumise, the Parti Socialiste, Europe Écologie Les Verts, and the Parti Communiste Français; the centrist bloc centred on Ensemble and its allies (MoDem, Horizons); and the far-right bloc, dominated by the Rassemblement National and its ally Union des Droites (UXD). With the far-right leading in first-round aggregate vote share (33.2\%), most pre-election polls forecast an RN majority \citep{rtlpred2024}. The second round produced the opposite outcome: the left obtained a narrow relative majority, driven by substantial centrist vote transfers, and by left-wing withdrawals in constituencies where the centrist candidate was better placed to block the far right.
This phenomenom of \emph{vote report} is illustrated in in Figure~\ref{fig:transfer-example}, which shows a schematic three-way second round: how a party's first-round voters redirect their votes toward each surviving bloc, and how the accumulated totals determine the final seat allocation.

This project investigates whether a mathematically principled model can anticipate such transfers, and therefore predict second-round outcomes, using only first-round vote shares and ideological party embeddings available prior to the vote. The model we develop distributes the votes of non-advancing parties across surviving candidates according to a softmax mechanism governed by ideological distance in a multi-dimensional embedding space derived from the Chapel Hill Expert Survey (CHES). Three aspects of the transfer problem shape the model's design:
\begin{enumerate}
    \item Transfers are partial. A centrist voter whose party does not advance does not necessarily switch to the ideologically nearest surviving candidate; they may split probabilistically across options, or abstain. A hard assignment rule such as "give all votes to the closest candidate", consistently mispredicts outcomes where two candidates are positioned near the midpoint between two ideological poles.
    \item The ideological space is multi-dimensional. The classic left–right axis conflates economic positions, cultural attitudes, nationalism, and anti-establishment sentiment, which are dimensions on which the 2024 French blocs diverged sharply and independently. A single axis is insufficient to capture the relevant distances.
    \item As incumbency matters, sitting members of parliament accrue local trust that may transcend party-ideological alignment. Therefore, a small incumbency bonus is given to candidates representing themself again after holding the seat for the last electoral period.
\end{enumerate}

\paragraph{Contributions.} This report makes four contributions. First, we propose a principled, parameter-efficient model of vote transfer grounded in a multi-dimensional ideological embedding (Section~\ref{sec:model}). Second, we provide an empirical calibration study across three electoral cycles (2017, 2022, 2024), demonstrating cross-election stability on hyperparameters of our model (Section~\ref{sec:calibration}). Third, a feature disaggregation analysis identifies which ideological dimensions drive predictive accuracy and demonstrates that electoral competition cannot be adequately captured by a simple left–right axis (Section~\ref{sec:features-split}). Fourth, a coalition disaggregation experiment assesses whether resolving union heterogeneity into constituent party vectors yields meaningful predictive gains (Section~\ref{sec:ug-split}).

\section{Data}
\label{sec:data}

\subsection{Electoral Results}

First and second round results at the \emph{circonscription} level are
obtained from the official French Ministry of the Interior open-data portal
for the 2017, 2022, and 2024 legislative elections \citep{minint2024}.
Only metropolitan France (mainland and Corsica) is retained; overseas
constituencies and constituencies representing the french abroad are excluded due to their substantially different political
dynamics and inconsistent data formatting. Constituencies in which a candidate
was elected outright in the first round are also excluded, as no second-round
transfer is observed.
After filtering, 489 constituencies remain for 2024, 534 for
2022, and 536 for 2017.

Each record contains: constituency code, candidate name, party affiliation
(\emph{nuance}), first-round vote count, second-round vote count (where
applicable), and an incumbent indicator (\emph{sortant}).

\subsection{Chapel Hill Expert Survey (CHES)}

Political party positions are drawn from the 1999--2024 Chapel Hill Expert
Survey dataset \citep{ches2024}. The reliability and cross-country validity
of CHES expert placements have been systematically evaluated by
\citet{bakker2015}, who confirm their stability across successive waves.
CHES collects expert ratings across a wide range of policy dimensions,
including left--right economic position, social (GAL--TAN) orientation,
attitudes toward European integration, immigration salience, climate salience,
internal party dissent, and several other indicators. For the 2017 election, we use the 2017 dataset. For the 2022 election, we use the 2019 dataset, and for the 2024 election, we use the 2024 dataset. Even if the datasets from 2017 and 2024 were published a few months after the elections happens, the alternative of using the 2012 and 2019 datasets for 2017 and 2024 wasn't a good option for this study, because of massive political diversification during the periods. From 2012 to 2017, the historical parties went throught a number of changes (names, program, political figures), and France also saw the rise of the center block, never truly considered before. Therefore, the 2012 dataset does not contain the data for the party that actually won the elections in 2017. Therefore, making predictions with it resulted in a signficant decreased accuracy (-15\%), as it failed to predict correctly this new party. For 2024, the 2019 data is also very old, and lack the response of the different parties towards the crisis that France had to face in the meantime (Covid19 pandemic, inflation, war in Ukraine), and perform therefore very poorly on the 2024 election. The 2022 election is also affected by that, but the data of 2019 is only 3 years apart, and perform therefore quite well on the 2022 election. 

French party codes are mapped to CHES entries via a
correspondence table. The majority of parties are directly present in the
CHES dataset under different names as in the French official datasets; several cases required non-trivial adjustments, which are
detailed in Appendix~\ref{app:party_map}.

Regarding the principal modelling choices: the Union of the Left is represented
by the mean of the CHES vectors of its constituent parties (FI, PS, EELV, PCF),
reflecting the fact that these parties not only coordinated their candidate
nominations but also presented a joint programme synthesising the platforms of
all coalition members. The centrist bloc, by contrast, is represented by
individual party vectors, as its member parties formed an electoral alliance
without adopting a common programme. The far-right alliance between RN and UXD
is handled by mapping UXD directly to the RN vector, reflecting the widespread
public perception of the two formations as ideologically indistinguishable.

\section{One-Dimensional Baseline}
\label{sec:baseline}

As a preliminary analysis, all parties are embedded on a single left--right
axis using the \emph{lrgen} feature from CHES, which ranks parties on a
general left--right ideological scale (expert combine economic ideology, cultural values, nationalism, social conservatism, attitudes toward redistribuition, and historical political tradition into a grade from 0 to 10, 0 being extreme left, and 10 extreme right). This baseline serves to establish a
lower bound on predictive accuracy and to quantify the limitations of
unidimensional party placement.

The classification of French parties according to CHES is reported in
Table~\ref{tab:lrgen-classification}.

\begin{table}[H]
\centering
\caption{Classification of french parties according to CHES (2024)}
\label{tab:lrgen-classification}
\scriptsize
\begin{tabular}{clc}
\toprule
Party & lrgen \\
\midrule
FI        & 0.82 \\
PCF       & 1.73 \\
UG        & 2.00 \\
LE/EELV   & 2.30 \\
PS        & 3.45 \\
MoDem     & 5.36 \\
Centre    & 5.98 \\
RE        & 6.27 \\
Horizons  & 6.60 \\
LR        & 7.73 \\
RN        & 8.82 \\
REC       & 9.73 \\
\bottomrule
\end{tabular}
\end{table}

Two baseline configurations are evaluated. A no-transfer baseline, in which
the first-round leading candidate is declared the winner without any vote
redistribution, achieves only 64.16\% accuracy on the 2024 elections, demonstrating the
substantive importance of modelling vote transfers.

A simple transfer rule is then applied, governed by three principles:

\begin{enumerate}\setlength\itemsep{2pt}
  \item A non-advancing party's votes are assigned entirely to the
        ideologically closest second-round candidate.
  \item If two candidates are equidistant, votes are split equally.
  \item An advancing party's votes are retained in full, under the assumption
        that first-round supporters do not switch candidates.
\end{enumerate}

Despite its simplicity, this baseline achieves approximately 71.24\%
constituency-level accuracy on 2024, confirming that ideological proximity
is a meaningful signal. The proximity model we adopt follows the Downsian
spatial tradition rather than the directional alternative of
\citet{rabinowitz1989}, in which voters prefer a candidate on the correct
side of an issue midpoint rather than the strictly nearest. In the French
runoff context, where voters face a constrained choice among qualifying
candidates, the proximity interpretation is more natural: the operative
question is which available option is least ideologically objectionable.
However, a 29\% error rate corresponds to approximately 134 mispredicted
constituencies out of 466, well above the tolerance required for electoral
forecasting. Notably, under this hard-assignment rule, all centrist votes are
redirected to the RN in any direct contest against the Union of the Left,
producing a predicted seat count of 165 for the far-right bloc.

This result is itself informative: even under a transfer scenario maximally
favourable to the far right, the model projects only 201 seats for the RN
bloc, far below the absolute majority of 289 seats forecast by several
pre-election analysts, suggesting that the structural configuration of the
electorate was less favourable to the far right than widely assumed.

Two structural limitations motivate the development of the full model:

(i) a single axis cannot capture the multi-dimensional nature of ideological
distances between parties;

(ii) the hard-assignment rule ignores partial transfers and is sensitive to
parties positioned near the midpoint between two second-round candidates.

\section{The Softmax Model}
\label{sec:model}

\subsection{Ideological Embedding via PCA}
\label{sec:pca}

Each French party $p$ is represented by a vector
$\mathbf{x}_p \in \mathbb{R}^{K}$ built from $K$ CHES features retained
after a feature selection from the CHES dataset. As the raw CHES data contains more than 45 numeric indicators, using them without
reduction risks over-weighting specific policy domains: attitudes toward
European integration, for instance, are captured by four separate variables,
while economic regulation positions are covered by seven, introducing an
implicit asymmetry in the ideological representation. Altought maybe some policies are more important than others in the eyes of the voter (and should therefore have more weight in the vectors), we chose to stick to a baseline where we tried to capture the most of the variance between the different parties, by focusing on capture most of the variance of the whole features. 
Our methodology was that pairs
of highly correlated features (Pearson $r > 0.85$) are identified and the
feature with lower variance is dropped from each pair.
The remaining features are standardised and submitted to Principal Component
Analysis.
We retain enough components to explain \textbf{95\% of the total variance},
yielding $K = \mathbf{15}$ dimensions for the CHES data of 2024.
The retained features, ranked by average absolute PCA loading, are listed
in Table~\ref{tab:pca}. This methodology is not perfect, but yielded better results on average (more or less 5\% more accuracy), that when it was not done

\begin{table}[H]
\centering
\caption{Retained CHES features (PCA-guided selection).}
\label{tab:pca}
\scriptsize
\begin{tabular}{clc}
\toprule
\textbf{\#} & \textbf{Feature} & \textbf{Importance} \\
\midrule
1  & \texttt{immigrate\_dissent}        & 0.235 \\
2  & \texttt{anti\_islam\_rhetoric}     & 0.233 \\
3  & \texttt{galtan\_blur}              & 0.228 \\
4  & \texttt{lrecon\_blur}              & 0.218 \\
5  & \texttt{deregulation}              & 0.218 \\
6  & \texttt{galtan\_dissent}           & 0.217 \\
7  & \texttt{antielite\_salience}       & 0.216 \\
8  & \texttt{regions}                   & 0.214 \\
9  & \texttt{religious\_principles}     & 0.209 \\
10 & \texttt{eu\_dissent}               & 0.208 \\
11 & \texttt{eu\_salience}              & 0.201 \\
12 & \texttt{lrecon\_salience}          & 0.199 \\
13 & \texttt{climate\_change\_salience} & 0.199 \\
14 & \texttt{eu\_blur}                  & 0.174 \\
15 & \texttt{lrecon\_dissent}           & 0.137 \\
\bottomrule
\end{tabular}
\end{table}

\noindent
The most discriminating features relate to \emph{internal party dissent}
(immigration, GAL-TAN, economic), anti-Islam rhetoric, anti-elite salience,
and regional autonomy, dimensions on which the three French blocs diverge
sharply, and which featured prominently in the 2024 campaign.
One notable observation is that the classic left--right economic position
(\texttt{lrecon}) does not appear among the top features: its variance is
partially absorbed by the corresponding dissent and salience indicators.

\subsection{Softmax Vote Transfer}
\label{sec:softmax}

The unidimensional hard-assignment rule evaluated in
Section~\ref{sec:baseline} yields unsatisfactory results, as it cannot
represent partial transfers across multiple second-round candidates. The
proposed model replaces this with a probabilistic softmax mechanism that
distributes votes from any non-advancing party across all second-round
candidates in proportion to ideological affinity, extending the discrete
choice framework of \citet{mcfadden1974} to the electoral transfer setting.
The use of multinomial logit-type models for multi-party electoral choice
has precedent in \citet{alvarez1998}, who show that such models outperform
simpler specifications in elections with three or more viable candidates ---
a setting directly analogous to the multi-bloc French second rounds studied
here.

Let constituency $c$ contain a set $\mathcal{R}_c$ of second-round candidates
and a set $\mathcal{S}_c$ of first-round parties that did \emph{not} advance.
For each $s \in \mathcal{S}_c$ with first-round total $v_s$, we compute a
score for each $r \in \mathcal{R}_c$:

\begin{equation}
  \label{eq:score}
  \mathrm{score}(s \to r) \;=\;
    -\delta \cdot \|\mathbf{x}_s - \mathbf{x}_r\|_2 + b_{s \to r},
\end{equation}

\noindent where $\delta \geq 0$ is a global \emph{distance sensitivity}
parameter and $b_{s\to r}$ is an incumbency bonus (Section~\ref{sec:bonus}). $\mathbf{x}_s$ and $\mathbf{x}_r$ are the vectors computed with the features of the CHES dataset. In our model, as we have selected 15 features, so $\mathbf{x}_s, \mathbf{x}_r \in \mathbb{R}^{15}$
The fraction of $v_s$ transferred to $r$ is:

\begin{equation}
  \label{eq:softmax}
  \pi(s \to r) \;=\;
    \frac{\exp\!\bigl(\mathrm{score}(s \to r)\bigr)}
         {\displaystyle\sum_{r' \in \mathcal{R}_c}
          \exp\!\bigl(\mathrm{score}(s \to r')\bigr)}.
\end{equation}

The predicted second-round total for candidate $r$ is:

\begin{equation}
  \label{eq:pred}
  \hat{V}_r \;=\; v_r^{(1)} \;+\;
    \sum_{s \in \mathcal{S}_c} v_s \cdot \pi(s \to r),
\end{equation}

\noindent where $v_r^{(1)}$ denotes the first-round votes of $r$ itself
(assumed fully retained).
The predicted winner is $\arg\max_r \hat{V}_r$.

A key advantage of the softmax formulation is that it naturally handles
second rounds with \emph{any} number of candidates (2, 3, 4, or even 5 in
some constituencies), since the denominator normalises across all options.
Identifying the optimal values of $\delta$ and $b_{s\to r}$ is non-trivial,
however, as small perturbations can substantially alter predicted outcomes.
Rather than fixing these parameters a priori, they are estimated via
calibration on past elections (Section~\ref{sec:calibration}).

\subsection{Incumbency Bonus}
\label{sec:bonus}

Sitting Members of Parliament (\emph{sortants}) seeking re-election receive
an additive bonus, reflecting the strong territorial rooting of French
legislative politics: an incumbent who has served a constituency for several
years is likely to have built a degree of personal trust that may transcend
strict party-ideological alignment, an advantage well documented across
electoral systems \citep{gelman1990}.

\begin{equation}
  \label{eq:bonus}
  b_{s \to r} =
  \begin{cases}
    c & \text{if } r \text{ is a sortant,} \\
    0              & \text{otherwise,}
  \end{cases}
\end{equation}

\noindent where $c \geq 0$ is the \emph{incumbency coefficient}.

\subsection{Model Summary}

The complete model has exactly \textbf{two free parameters}:

\begin{center}
\small
\begin{tabular}{ccp{3.2cm}}
\toprule
Param. & Domain & Interpretation \\
\midrule
$\delta$ & $[0,\,5.0]$ & Sensitivity to ideological distance \\
$c$      & $[0,\,2.5]$  & Incumbency advantage multiplier \\
\bottomrule
\end{tabular}
\end{center}

Both parameters are shared across all parties and constituencies, making the
model parsimonious and interpretable. Additional parameters could in principle
be introduced to capture candidate-specific notoriety effects, for
instance, the electoral advantage accrued by former ministers or prominent
public figures. However, any such extension would require subjective
judgements about which candidates merit a bonus, introducing an undesirable
degree of researcher discretion. The two-parameter specification is therefore
retained throughout. Having two parameter is also useful for parameter optimization, as a 2d heatmap is easier to read than a 1d heatmap, as it allows us to differciate the different effects represented by these parameters. The range of the parameters have been decided throught trial an error, we saw that the best parameter were always in this range. We expanded sometimes the range to $\delta = 10$, to see if the accuracy was rapidly decreasing when $\delta \in [5, 10]$, but the best parameter were always found within the initial range.

\section{Calibration and Validation}
\label{sec:calibration}

\subsection{Grid Search}

We calibrate $(\delta, c)$ by exhaustive grid search over:
\[
  \delta \in \{0.00,\,0.01,\,\ldots,\,5.00\},\quad
  c \in \{0.00,\,0.01,\,\ldots,\,2.50\},
\]
yielding more than 100{,}000 candidate pairs. These numbers have been refined on larger grid search, to optimize the interval where the accuracy is the highest in general (see Figures~\ref{fig:accuracy_matrix_2017} ~\ref{fig:accuracy_matrix_2022} ~\ref{fig:accuracy_matrix_2024})
For each pair, we compute constituency-level accuracy on the 2022 and 2017
elections separately, with the objective of identifying parameters suitable
for out-of-sample prediction of the 2024 election. The grid search is
parallelised across CPU cores to reduce computation time, and takes around 3 minutes to complete with a AMD Ryzen 7 3700X.

\subsection{Accuracy Results}

Table~\ref{tab:accuracy} reports the best constituency-level accuracy achieved
for each election when the model is optimised directly on that election's data.
These in-sample results serve as an upper bound; the out-of-sample 2024
predictions are discussed in Subsection~\ref{subsec:results}.

\begin{table}[H]
\centering
\caption{Best Constituency-level accuracy achieved across elections.}
\label{tab:accuracy}
\small
\begin{tabular}{lccc}
\toprule
\textbf{Election} & \textbf{Role} & \textbf{$N$} & \textbf{Acc.\ (\%)} \\
\midrule
2017 & Calibration & 536 & 86.94 \\
2022 & Calibration & 534 & 87.45 \\
2024 & Evaluation  & 489 & \textbf{91.85\%} \\
\bottomrule
\end{tabular}
\end{table}

\noindent
The optimal parameters from the 2017 calibration are $\delta^* = 3.17$ and
$c^* = 0.33$; those from 2022 are $\delta^* = 3.23$ and $c^* = 0.05$.
In the both cases, multiple parameter pairs yield identical peak accuracy; the
median pair is reported. Across both elections, the accuracy landscape forms
a broad plateau rather than a sharp peak, indicating robustness to small
perturbations in the parameter values. Nevertheless, the optimal incumbency
coefficient $c^*$ differs substantially between the two elections, reflecting differences in political context:
2017 corresponded to a period of relative institutional stability, whereas 2022
followed the disruptions of the COVID-19 pandemic and the ensuing inflation
crisis, and therefore the already elected people were subject to move, as people wanted some changes.

\subsection{Accuracy Landscape}

\begin{figure}[H]
  \centering
  \includegraphics[width=\linewidth]{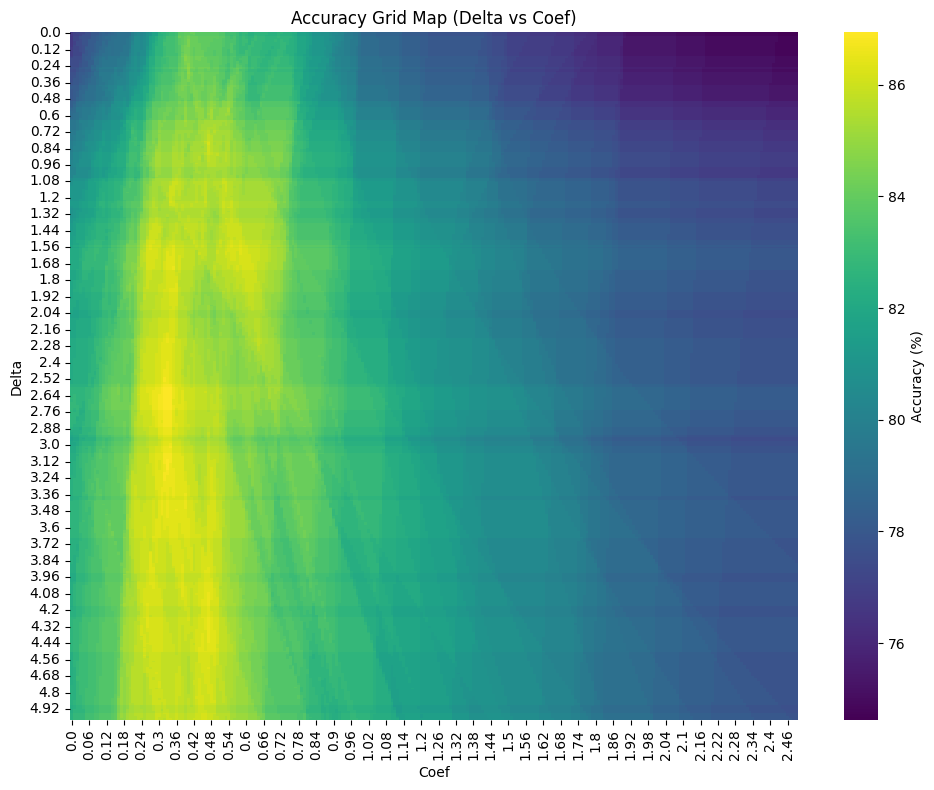}
  \caption{Accuracy heat map over the $(\delta,\,c)$ grid (2017 election).
  Brighter colours indicate higher accuracy.}
  \label{fig:accuracy_matrix_2017}
\end{figure}

\begin{figure}[H]
  \centering
  \includegraphics[width=\linewidth]{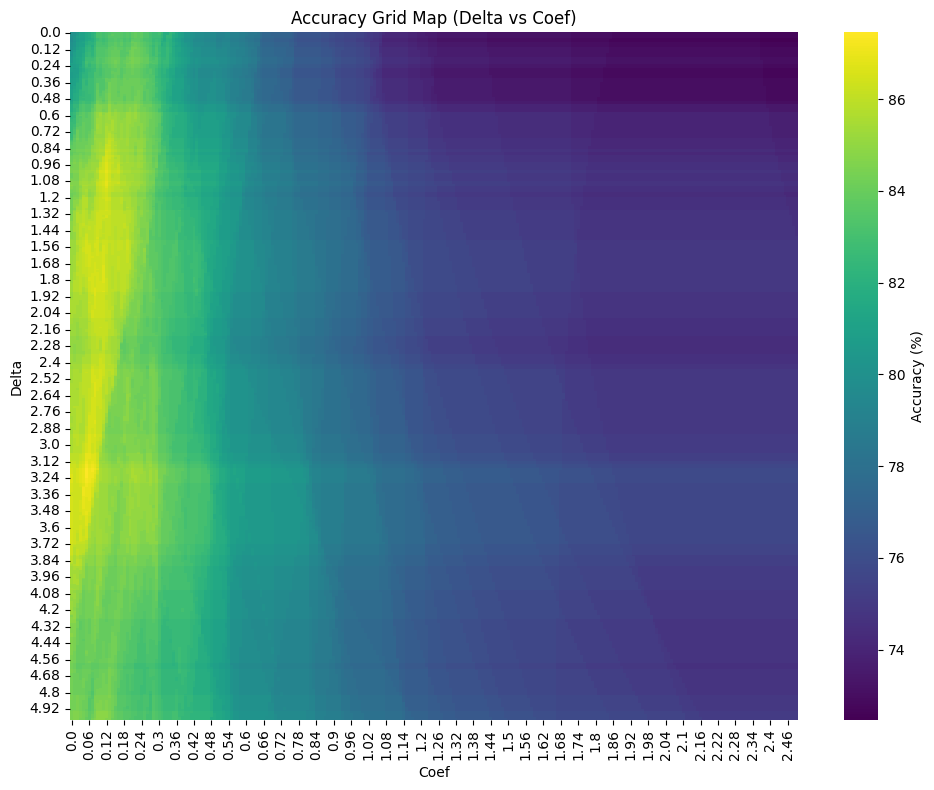}
  \caption{Accuracy heat map over the $(\delta,\,c)$ grid (2022 election).
  Brighter colours indicate higher accuracy.}
  \label{fig:accuracy_matrix_2022}
\end{figure}

Figure~\ref{fig:accuracy_matrix_2017} and Figure~\ref{fig:accuracy_matrix_2022}
display the accuracy heat map over the $(\delta, c)$ grid for 2017 and 2022
respectively.
In both cases, the optimal region forms a ridge rather than a narrow peak,
confirming that the calibrated values generalise well out of sample.
While the optimal range for $\delta$ is relatively stable across elections
($\delta \in [2.5, 4.0]$), the optimal $c$ varies considerably. Setting
$c = 0$ on the 2017 data reduces accuracy from 86.94\% to approximately
77.34\%, illustrating the sensitivity of predictions to the incumbency
specification and the extent to which $c$ encodes election-specific political
conditions.

\subsection{Results of the model on the 2024 election}
\label{subsec:results}

Inspection of the accuracy landscape across both calibration years suggests
that strong performance is consistently achieved for
$\delta \in [2.5, 4.0]$ and $c \in [0.06, 1.0]$. Taking the midpoint of the range for delta yields $\delta = 3.25$. For $c$, we saw it was quite hard to determine, and as the 2024 elections were called shortly and unexpedetly, we can assume a political climate were the person already elected in 2022 were likely to be reelected, as people initially elected them for 5 years. Therefore, we chose $c = 0.7$, which achieves
90.34\% accuracy, close to the in-sample optimal of 91.85\% obtained at
$\delta = 3.3$ and $c = 0.78$. These consensus parameters
($\delta = 3.25$, $c = 0.7$) are adopted for all subsequent analyses, as
they represent the best achievable performance under a prospectively valid
calibration strategy and yield consistent results across both the 2017 and
2022 elections.

\begin{figure}[H]
  \centering
  \includegraphics[width=\linewidth]{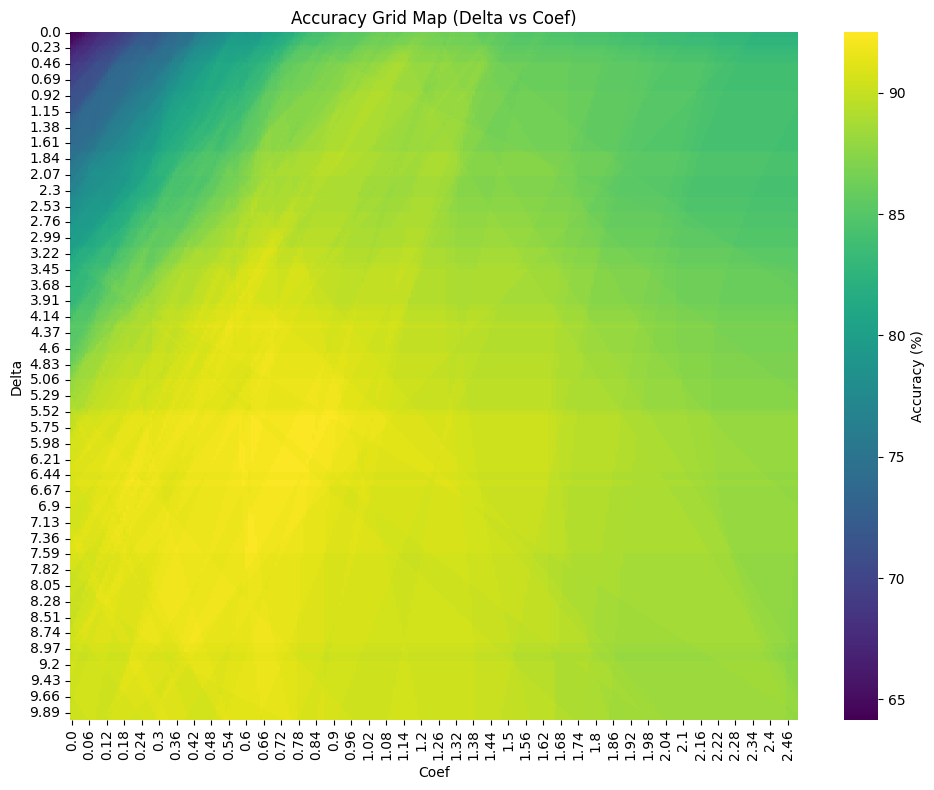}
  \caption{Accuracy heat map over the $(\delta,\,c)$ grid (2024 election).
  Brighter colours indicate higher accuracy.}
  \label{fig:accuracy_matrix_2024}
\end{figure}

Figure~\ref{fig:accuracy_matrix_2024} shows the accuracy landscape for 2024.
The optimal incumbency coefficient is substantially higher in this election
than in previous cycles, consistent with the circumstances of the vote: the
2024 election was called abruptly following the dissolution of an assembly
elected only two years prior. In many constituencies, voters re-elected the
sitting member of parliament as a signal of confidence in a locally known representative.
Notably, even with $c = 0$, the model achieves approximately 85\% accuracy,
confirming that the incumbency bonus is a non-trivial component and plays a
meaningful role in pushing accuracy above the 90\% threshold. We chose to expand the calibration above $\delta = 5$, to $\delta = 10$, because we saw in the original one that the model was likely to yield consistent results even after the initial treshold.

\subsection{Per-Party Breakdown (2024)}

Table~\ref{tab:per_party} reports per-party accuracy on the 2024 evaluation
set using the calibrated parameters.

\begin{table}[H]
\centering
\caption{Per-party accuracy, 2024 election.}
\label{tab:per_party}
\scriptsize
\begin{tabular}{lcccc}
\toprule
\textbf{Party} & \textbf{Predicted} & \textbf{Correct} & \textbf{Incorrect} & \textbf{Real} \\
\midrule
RN (+ allies)  & 93 & 78  & 15 & 86  \\
UG (FI, EELV, PS, PCF) & 146 & 132 & 14 & 141 \\
ENS (+ allies) & 150 & 137 & 13 & 139  \\
LR             & 33  & 33  & 0 & 38 \\
DVD            & 15  & 15  & 0 & 22 \\
UXD            & 9   & 7   & 2 & 16 \\
DVG            & 6  & 6  & 0 & 6 \\
HOR            & 6   & 6   & 0 & 6 \\
REG            & 4   & 3   & 1 & 3 \\
DVC            & 2   & 2   & 0 & 4 \\
ECO            & 1   & 1   & 0 & 1 \\
UDI            & 1   & 1   & 0 & 3 \\

\midrule
\textbf{Overall} & 466 & 421 & 45 & 466 \\
\bottomrule
\end{tabular}
\end{table}

\noindent
For reference, DVD (Divers Droite), DVC (Divers Centre), and DVG (Divers
Gauche) designate candidates without formal party affiliation but whose
programmes are proximate to the corresponding political family. These are the labels attributed in the data of the french government.
Small centrist and right-wing parties (LR, DVD, HOR, DVC, DVG, UDI) are
predicted with high accuracy, as their candidates typically face ideologically
unambiguous transfer situations. The moderate misprediction rate for UXD
is expected: as a newly formed formation without a dedicated CHES entry, UXD
is mapped to the RN vector, which underestimates its ideological proximity
to centrist candidates.

For the three principal blocs, the error rate is approximately 7--8\%, which
compares favourably to the pre-election survey predictions. Most notably,
the model assigns 157 seats to the Center (classified as ENS + allies, as well as HOR and UDI), whereas most major
polling organisations at the time forecast fewer than 100 seats for this
coalition \citep{rtlpred2024}, for it to win 148 seats in the end.

Examining the systematic errors reveals that the model predicts ENS in place
of UG in 6 cases, UG in place of UXD in 5 cases (a predictable consequence
of mapping UXD to the RN vector), and UG in place of RN in 5 cases.
The model also over-predicts RN in direct contests against right-wing parties
(LR or DVD), accounting for 7 additional errors. The frequency with which UG
features in the error cases motivates the coalition disaggregation experiment
of Section~\ref{sec:ug-split}: as shown in Table~\ref{tab:lrgen-classification},
the PS is positioned nearly as close to MoDem on the left--right scale as it
is to FI, implying that centrist voters would have greater incentive to
transfer to the PS in a PS--RN contest than to FI in a FI--RN contest.

\subsection{Seat Distribution and Aggregate Bias}

\begin{figure}[H]
  \centering
  \includegraphics[width=\linewidth]{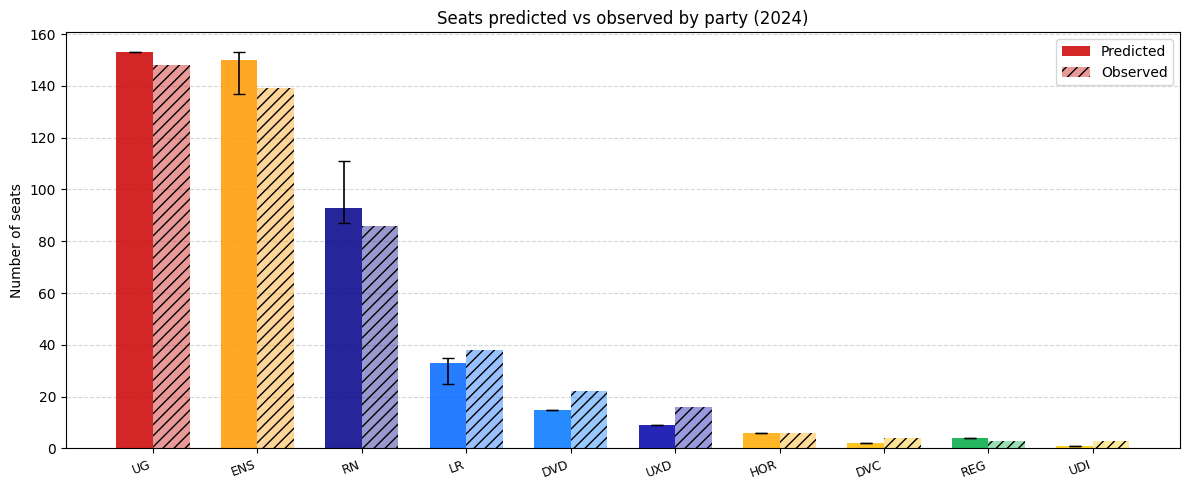}
  \caption{Predicted vs.\ actual seat totals by bloc (2024).}
  \label{fig:seat_distribution}
\end{figure}

Even where the model mispredicts individual constituencies, many errors cancel
in aggregate: ENS is over-predicted in some seats and under-predicted in
others, so the total seat count per bloc is more accurate than
constituency-level accuracy alone would suggest.
Figure~\ref{fig:seat_distribution} compares predicted and actual seat totals.

\subsection{Interpretation of the different elections}

Across the three electoral cycles, the distance sensitivity parameter $\delta$ proves remarkably stable: its optimal value shifts little from one election to the next, suggesting that the role of ideological proximity in governing vote transfers is a structural feature of the French system rather than a contingent property of any particular election. The incumbency coefficient $c$, by contrast, varies more across cycles, though it consistently remains small, in the range $[0.05,0.5]$. The 2024 election stands out as the cycle with the highest estimated $c$: this is consistent with the unusual circumstances of a snap election called after a presidential dissolution of the Assembly, in which incumbents may have benefited from heightened name recognition and local visibility at a moment when parties had little time to reorganise their candidate slates. Despite this variability, the heatmaps show that the incumbency bonus improves accuracy in all three elections. The heatmaps further show that setting both $\delta$ and $c$ to zero, thereby distributing transfer votes uniformly across all surviving candidates, causes a sharp drop in accuracy, confirming that ideological distance carries genuine predictive signal. The consistently small magnitude of $c$ relative to $\delta$ also clarifies its functional role: the incumbency bonus does not redirect large vote shares, but provides a small systematic nudge that proves decisive in the many constituencies where the margin between candidates is narrow.

A related pattern emerges when predictions are disaggregated by winning margin. The model predicts correctly in nearly all constituencies where the second-round winner prevails by more than ten percentage points, and errors concentrate almost exclusively in tight races. This is a natural consequence of the model's design: because the softmax mechanism distributes transfers continuously rather than assigning them wholesale, no candidate is left with zero transferred votes, and large leads are structurally robust to the precise allocation. Geographically, wide-margin constituencies tend to coincide with dense urban cores, Paris, Lyon, Marseille, where the political landscape is more homogeneous and bloc boundaries more clearly drawn. Narrow-margin constituencies, and thus the bulk of mispredictions, are concentrated in peri-urban and rural areas, where local factors, candidate notoriety, and within-bloc heterogeneity play a larger role than the model's ideological embeddings can capture.

One modelling choice deserves explicit justification. We considered augmenting the transfer model with constituency-level socioeconomic covariates, for instance, giving a systematic bonus to left-wing candidates in lower-income areas. We ultimately chose not to do so, for two reasons. First, the first-round vote share of each party already functions as a proxy for the socioeconomic composition of the constituency: voters reveal their preferences in the first round, and those preferences already incorporate their economic situation and interests. Adding an explicit socioeconomic bonus would therefore risk double-counting information already present in the first-round totals. Second, in the constituencies where socioeconomic conditions are most extreme, second-round contests frequently oppose two candidates both classified as left-wing but differing in radicalism or programme, a distinction the socioeconomic index cannot resolve. For both reasons, we retain a model that conditions exclusively on first-round vote shares and ideological party embeddings.

Further experiments also show that the best result is always obtained with the nearest data from CHES, regardless of the election. However, the accuracy change is only a few points each time, meaning previous political placement stays important for an election, even 5 or 10 years later. This is a mitigating result toward \cite{adams2009}, which claims that voters take one electoral term before considering the policy shift. However, in the recent years, the political spectrum in France has evolved at a higher speed that at the time of the experiments of the paper. In the last 15 years, it has been recuring that a newly founded party manages to obtain significant results at the next election (for example, Emmanuel Macron founded his party only 2 years before winning the presidential election in 2017). The historical french parties also changed structuraly, or became medium-sized parties (UMP, an historical party that won the 2007 presidential election, transformed into LR, which now holds less than 10\% of the votes, and the PS went to winning the presidential election in 2012 to having a score of less than 7\% in 2017), due to the rise of newly created parties. Moreover, there is also multiple intermediate elections between two terms, giving the time for the voters to adjust to the new policy shift of the different parties. Nevertheless, the old policy still plays a role, and voters tend to align partially on it, since using an old dataset on any election almost always guarantee an accuracy above 85\% with an hyperparameter calibration.

\section{Features Observation}
\label{sec:features-split}

\subsection{Motivation}

The PCA-based feature selection described in Section~\ref{sec:pca} identifies
the dimensions along which French parties diverge most in the CHES embedding
space. However, explaining variance between parties is not equivalent to
explaining predictive accuracy for vote transfers: a dimension that sharply
separates party positions may carry little electoral weight if it did not
feature prominently in the campaign. To assess the predictive contribution
of each retained feature, the model is re-run using each of the 15 selected
features in isolation, with fixed parameters $\delta = 3.25$ and $c = 0.7$.

\subsection{Accuracy depending on the feature}

\begin{table}[H]
\centering
\caption{Impact of the features on accuracy (2024).}
\label{tab:impact_features_accuracy}
\small
\begin{tabular}{lcc}
\toprule
\textbf{Feature} & \textbf{Acc.\ (\%)} & \textbf{$\Delta$} \\
\midrule
Base (all features together)  & 90.34 & --- \\
immigrate dissent & 89.70 & -0.64 \\
anti islam rhetoric & 91.20 & +0.86 \\
galtan blur & 89.06 & -1.28 \\
lrecon blur & 88.41 & -1.93 \\
deregulation & 76.39 & -13.95 \\
galtan dissent & 86.05 & -4.29 \\
antielite salience & 87.55 & -2.79 \\
regions & 82.83 & -7.51 \\
religious principles & 81.76 & -8.58 \\
eu dissent & 80.69 & -9.65 \\
eu salience & 83.26 & -7.08 \\
lrecon salience & 87.34 & -3.00 \\
climate change salience & 80.47 & -9.87 \\
eu blur & 89.09 & -1.25 \\
lrecon dissent & 85.62 & -4.62 \\

\bottomrule
\end{tabular}
\end{table}


Below is a summary of the different features:

\begin{itemize}\setlength\itemsep{1pt}

\item \textbf{Immigrate dissent} measures the degree of internal disagreement
within a party regarding immigration policy. A higher level of dissent indicates
that party elites hold divergent positions on immigration-related issues.

\item \textbf{Anti-Islam rhetoric} captures the extent to which a party employs
rhetoric critical of Islam or frames Islam as incompatible with national values
or institutions. 

\item \textbf{GAL--TAN blur} measures ambiguity or lack of clarity in a party’s
position on the Green/Alternative/Libertarian versus Traditional/Authoritarian/Nationalist
dimension. 

\item \textbf{LRECON blur} captures ambiguity in a party’s economic left--right
position. Parties with blurred economic positioning may attract more heterogeneous
electorates, affecting transfer patterns between rounds.

\item \textbf{Deregulation} measures support for reducing market regulations and
state intervention in the economy. 

\item \textbf{GAL-TAN dissent} reflects internal disagreement within a party on
sociocultural issues such as immigration, authority, nationalism, and traditional
values.

\item \textbf{Anti-elite salience} measures the importance a party places on
anti-establishment or anti-elite discourse in its political messaging.

\item \textbf{Regions} captures support for regional autonomy, decentralisation,
or the political importance of territorial identities.

\item \textbf{Religious principles} measures the extent to which a party supports
the role of religious values or principles in politics and public life.

\item \textbf{EU dissent} reflects internal disagreement within a party regarding
European integration and the European Union.

\item \textbf{EU salience} measures the importance of European integration issues
within a party’s political agenda.

\item \textbf{LRECON salience} captures the importance of economic left--right
issues in a party’s platform and political competition.

\item \textbf{Climate change salience} measures the degree to which climate and
environmental issues are emphasised by a party.

\item \textbf{EU blur} captures ambiguity or inconsistency in a party’s position
toward the European Union and European integration. 

\item \textbf{LRECON dissent} measures internal disagreement within a party on
economic left--right issues such as taxation, redistribution, and market
regulation.

\end{itemize}

One feature stands out: \texttt{anti\_islam\_rhetoric}, interpretable as a
proxy for immigration attitudes, achieves 91.20\% accuracy when used in
isolation, marginally exceeding the full 15-feature model.
This result must be interpreted with caution, since
it is obtained by fitting to the 2024 outcome data and are therefore not
generalisable to future elections without additional validation. The accuracy
gain is likely driven by the particular salience of immigration as a campaign
issue in 2024, and may not replicate in cycles where the political agenda is
structured differently.

Several other features, notably \texttt{deregulation},
\texttt{climate\_change\_salience}, and \texttt{eu\_salience}, yield
substantially lower accuracy when used alone. This is consistent with either
of two interpretations: the corresponding policy dimensions may have been
largely absent from the 2024 campaign discourse, or the parties' positions
on these dimensions may be insufficiently differentiated to generate a
reliable transfer signal.

\section{Disaggregating the Unions}
\label{sec:ug-split}

\subsection{Motivation}

In the baseline model, all UG candidates are assigned the same ideological
vector: the mean of the CHES vectors of FI, PS, EELV, and PCF. We chose a non weighted mean, as the parties were contributing equaly to the common program. Moreover as there is only 4 parties, the median will exclude two parties from the computation, so it was likely to introduce some error. During the developement, we tested mean and median multiple times, and the mean was more stable in the results, altought in the final model it yields the same accuracy.
This aggregation may introduce systematic error describe as follow. A FI candidate, positioned
at the far left of the ideological spectrum, will attract a substantially
different flow of centrist transfers than a PS candidate positioned near the
centre-left. Assigning both the same coalition-mean vector masks this
heterogeneity. The same reasoning applies to the centrist coalition, which
brings together UDI, Horizons, and Ensemble.

\subsection{Implementation}

For 2024, the ``UG'' label on official candidate registration forms does not
distinguish between the constituent parties of the alliance. Individual party
affiliations (FI, PS, EELV, PCF) are recovered by cross-referencing candidate
names and constituency codes against scraped HTML data from the official party
websites. Each recovered candidate is then assigned the CHES vector of their
declared party rather than the coalition mean. The constituent parties of the
centrist coalition are already individually identified in the original data.

\subsection{Results}

Table~\ref{tab:ug_split} compares accuracy with and without UG and centrist
disaggregation.

\begin{table}[H]
\centering
\caption{Effect of UG disaggregation on accuracy (2024).}
\label{tab:ug_split}
\small
\begin{tabular}{lcc}
\toprule
\textbf{Model variant} & \textbf{Acc.\ (\%)} & \textbf{$\Delta$} \\
\midrule
Aggregated UG and Center  & 90.34 & ----- \\
Aggregated UG only & 90.34 & +0 pp \\
Aggregated Center only & 89.48 & -0.86 pp \\
Disaggregated UG and Center & 89.48 & -0.86 pp \\

\bottomrule
\end{tabular}
\end{table}

\noindent
Two findings emerge from this experiment. First, disaggregating the centrist
coalition produces no accuracy gain: since each centrist party fielded at most
one candidate per constituency and the member parties are positioned closely
together in the ideological space, their disaggregation does not meaningfully
alter predicted transfer flows.

Second, disaggregating the Union of the Left into its constituent parties
produces a slight accuracy decline. Resolving UG into individual party vectors
causes the model to assign lower transfer probabilities to FI candidates in
contests where centrist voters are the pivotal group, a theoretically
intuitive result given the large ideological distance between FI and MoDem.
At the same time, the model tends to confuse PS candidates with centrist
candidates, as their ideological vectors are sufficiently proximate that
transfer flows from other parties split more evenly between them, producing
more frequent prediction errors in constituencies where the race is close.
The net effect is a marginal loss in aggregate accuracy, suggesting that the
coalition mean provides a more robust ideological representation for prediction
purposes than individual party vectors, at least at the current level of
model complexity.

\section{Discussion}
\label{sec:discussion}

\paragraph{Relation to prior work.}
The closest empirical antecedent of this project is \citet{fauvelle2008}, who
study the relationship between first- and second-round vote flows in French
legislative elections and confirm that first-round results systematically
predict runoff outcomes. The present project shares that motivation but differs
in three respects: transfers are grounded in multi-dimensional CHES
ideological embeddings rather than derived from aggregate vote shares alone;
a probabilistic softmax mechanism replaces regression-based flow estimation;
and predictive accuracy is evaluated out of sample across three electoral
cycles, including the structurally novel three-bloc contest of 2024, which
\citeauthor{fauvelle2008} could not have considered.

\paragraph{Strengths.}
The model is parsimonious (two free parameters), interpretable
(distance-based transfer with incumbency correction), and transferable across
electoral cycles, as evidenced by its consistent performance on 2017 and 2022
data. Its accuracy increases monotonically from 2017 to 2024, which may
reflect a progressive crystallisation of the French party system around three
coherent ideological blocs, a structure that the CHES embeddings appear to
capture increasingly well over time. A practical advantage of the model is
its reliance on publicly available data: official first-round results are
typically published within 24--48 hours of the vote, enabling prospective
second-round predictions before the inter-round campaign period closes, as the code is designed to be quick to adapt to any new dataset format, and the model runs in a few seconds. The
model's primary objective, maximising the number of correctly predicted
constituency winners, is well suited to seat projection, the key quantity
for assessing government formation scenarios.

\paragraph{CHES dataset.}
The model depends heavily on the CHES dataset for its ideological embeddings,
and is therefore sensitive to how accurately and promptly CHES data reflects
the current positions of political parties. The French party landscape can
evolve rapidly over a parliamentary term: parties may be founded, dissolved,
renamed, or undergo significant programmatic shifts. Given that CHES data
collection cycles span six to eight years, there is a real risk that the
available wave does not adequately represent the parties competing in a given
election. The present study illustrates this directly: applying the 2017 CHES
wave to the 2022 election,in which several major parties had changed names
and programmes substantially, reduces accuracy to approximately 85\%,
compared to the 88\% achieved using the 2024 wave retrospectively. Furthermore,
CHES data are collected through expert surveys, and inevitably carry the biases
inherent in that mode of data collection, including judgements about which
policy dimensions to include and how to present them to respondents.

\paragraph{Metropolitan France.}
The analysis is restricted to metropolitan France, excluding approximately 40
constituencies (around 7\% of all constituencies) in overseas territories and
those representing French citizens abroad. These seats can be pivotal in
determining whether a party achieves an absolute or merely a relative majority.
Their exclusion is motivated by the substantially different political dynamics
of these constituencies relative to metropolitan France; however, any
application of the model to full-assembly seat projections must account for
this gap. Empirically, the overseas tend to favour the left parties, and the abroad constituencies tend to favour
the presidential party: in 2024, the Union of the Left obtained approximately
20 such seats while the centrist bloc obtained around 10, a differential that
partially explains the discrepancy between our near-parity prediction for these
two blocs and the actual outcome.

\paragraph{PCA features.}
The PCA-based feature selection produces an unbiased representation of
ideological party differences, but selects dimensions on the basis of
explained variance in the CHES embedding space, not on their predictive
relevance for vote transfers. The feature analysis in
Section~\ref{sec:features-split} shows that using features in isolation
almost universally reduces accuracy, yet some single features outperform the
full model on the 2024 data. A more targeted feature selection guided by
campaign salience would likely improve accuracy, but would require a priori
judgements about the political agenda that risk introducing researcher bias
and would reduce the model's generalisability to future elections. We also chose to retain 95\% of the variance, which is an arbitrary number, yielding an arbitrary number of features depending on the CHES dataset (for example, the 2017 dataset has 16 features explaining above 95\% of the variance whereas the 2024 dataset only has 15 features explaining the same amount). The other problem on using PCA analysis, and correlation to rule out some features is that maybe two different features share the same distribution accross parties (parties get the same grade by the expert on these features), but are in fact very different and should be consider separate. The primary goal of doing a PCA was to retain one feature per "group" of features, but it has this disavantage of maybe overlapping two or more groups of features into one unique. The initial goal of the PCA was also to extract the "primary" component of a group, but can sometimes extract a proxy component that has more variance. For example \textit{anti-islam-rethoric}, which is a proxy for \textit{immigration}. Altought we run some expriment to confirm our strategy of doing a PCA analysis and retain variance, it is also worth mentionning that running the model with every PCA features only decrease accuracy by a few points (between 3\% and 5\% depending on the analysed election). One thing that could also be improved in this regard would be to optimize the features selection based on previous elections instead of doing a PCA analysis, in the same way we optimized the hyperparameters. 

\paragraph{Candidate heterogeneity.}
Candidates are represented solely by their party affiliation, yet individual
candidates within the same party may hold meaningfully different positions.
This is visible even within Ensemble, whose candidates range from centre-right
to centre-left on several dimensions. Intra-party heterogeneity also manifests
at the voting level: it is not uncommon for elected members of parliament to vote against
their party's line in the assembly for ideological or strategic reasons.
Additionally, candidate-specific factors, the notoriety of a former
president or minister, local visibility built over several terms, or the
reputational consequences of recent political scandals, affect individual
electoral performance in ways the model does not capture. While the incumbency
bonus partially proxies for these effects, it cannot account for the full
range of candidate-level variation.

\paragraph{Abstention.}
The model implicitly assumes a stable electorate between rounds. In 2024, the
abstention rate remained nearly constant (less than 0.1 percentage points
difference between rounds), making this assumption reasonable. In 2022,
however, turnout differed by more than 1.3 percentage points across the two
rounds, representing over half a million additional votes cast in the second
round relative to the first. Differential mobilisation can substantially alter
constituency-level outcomes, particularly in close races, and constitutes a
source of prediction error the present model does not address.

\paragraph{Training objective.}
The model is optimised to maximise the fraction of correctly predicted
constituency winners. An alternative objective, minimising the mean
absolute error in predicted vote shares, would likely yield different
parameter estimates and potentially different predictions in marginal
constituencies. The current approach prioritises seat projection accuracy over
vote-share estimation; researchers interested in the latter quantity may wish
to re-calibrate the model accordingly.

\paragraph{Uniform transfer assumption.}
The model assumes all voters of a given party transfer according to the same
ideological logic. In practice, transfers may be heterogeneous within a party:
local factors such as candidate visibility, explicit voting directives issued
by party leaders between rounds, or idiosyncratic constituency dynamics can
generate systematic deviations from the national-level ideological signal.

\section{Conclusion}
\label{sec:conclusion}

We have developed a softmax vote-transfer model for French legislative
elections grounded in multi-dimensional ideological embeddings from the CHES
dataset.
After reducing the 45-feature CHES space to 15 dimensions via PCA (retaining
95\% of variance), transfers are allocated via an exponential scoring rule
penalising ideological distance, with an incumbency bonus that preserves
interpretability.

Calibrated on 2017 and 2022 elections, the model achieves \textbf{90.34\%}
constituency-level accuracy on 2024, up from $\sim$71.24\% for the
one-dimensional baseline.
Disaggregating the Union of the Left or the centrist coalition into their
constituent parties does not improve predictive accuracy under the current
calibration, and can modestly reduce it, suggesting that coalition-level
ideological vectors provide sufficient resolution for the purposes of
constituency-level winner prediction.

The model's parsimonious design, strong out-of-sample performance, and
transparent interpretability establish it as a useful baseline for electoral
forecasting in two-round systems, and provide concrete evidence that
ideological distance is the dominant driver of inter-round vote transfers
in French legislative elections.

\balance    

\end{multicols*}

\newpage
\bibliographystyle{plain}

\appendix

\section{Party Mapping}
\label{app:party_map}

\begin{table}[H]
\centering
\caption{French electoral nuance codes and CHES mapping.}
\label{tab:party_map}
\small
\begin{tabular}{p{3.5cm} p{3.5cm} p{3.5cm}}
\toprule
\textbf{Nuance code(s)} & \textbf{Description} & \textbf{CHES entry} \\
\midrule
DXG, EXG, RDG & Far / radical left & PCF \\
FI, PCF, NUP, UG, ECO, EELV, DVG, PS & Left / centre-left & Mean(PCF, FI, EELV, PS) \\
UDI, DVC, HOZ, HOR, ENS, REG & Centre & Mean(ENS, MoDem, HOR) \\
LR, DVD & Right & LR \\
UXD, RN, EXD, REC, DSV & Far right & RN, REC \\
\bottomrule
\end{tabular}
\end{table}

\section{Full Accuracy Matrix}
\label{app:accuracy_matrix}

\begin{figure}[H]
  \centering
  \includegraphics[width=0.85\textwidth]{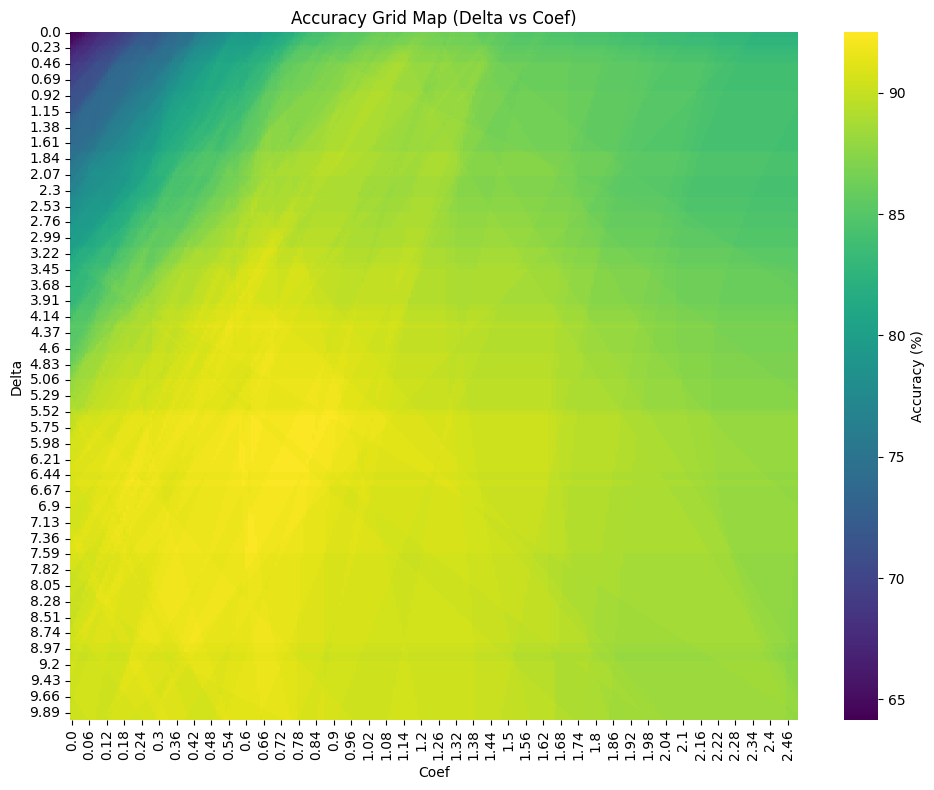}
  \caption{Full accuracy matrix for the 2024 election over the grid
  $\delta \in [0.00,\,10.00]$, $c \in [0.00,\,2.50]$.}
  \label{fig:accuracy_full}
\end{figure}

\section{Correct and predicted results}
\label{app:correct_france}

\begin{figure}[H]
  \centering
  \includegraphics[width=0.85\textwidth]{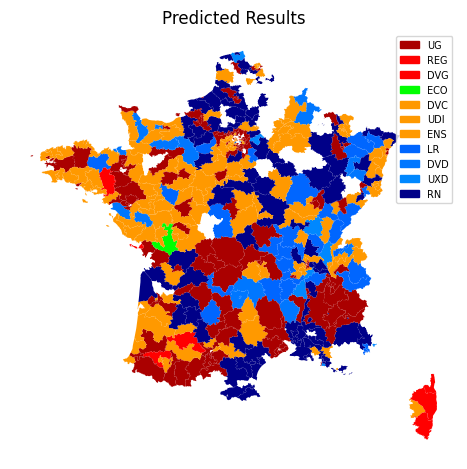}
  \caption{France map with the predicted results by the model}
  \label{fig:france_predicted}
\end{figure}
\begin{figure}[H]
  \centering
  \includegraphics[width=0.85\textwidth]{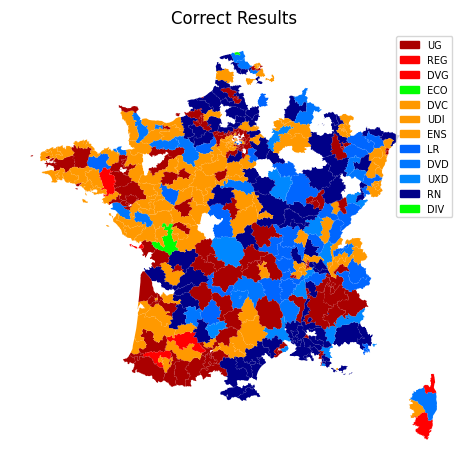}
  \caption{France map with the correct results}
  \label{fig:france_correct}
\end{figure}

\section{Worked example}
\label{app:example}

Consider a constituency where the second round opposes ENS and RN, with FI
and PS having substantial first-round shares.
With $\delta = 3.15$, $c = 0.7$, and the 15-dimensional CHES vectors:

\begin{align*}
  \mathrm{score}(\mathrm{UG} \to \mathrm{ENS})
    &= -3.15 \times \|\mathbf{x}_\mathrm{FI} - \mathbf{x}_\mathrm{ENS}\|_2
    \approx 0.10, \\
  \mathrm{score}(\mathrm{UG} \to \mathrm{RN})
    &= -3.15 \times \|\mathbf{x}_\mathrm{FI} - \mathbf{x}_\mathrm{RN}\|_2
    \approx 0.07.
\end{align*}

The softmax assigns, e.g.,
$\pi(\mathrm{UG} \to \mathrm{ENS}) \approx 0.76$ and
$\pi(\mathrm{UG} \to \mathrm{RN}) \approx 0.23$.
If we replace UG by PS, being closer to ENS in the ideological space, the model will assigns a larger fraction:
$\pi(\mathrm{PS} \to \mathrm{ENS}) \approx 0.83$, illustrating
why disaggregating UG into specific parties such as PS changes predictions in centrist-adjacent
constituencies.

\end{document}